\documentclass[preprint,aps]{revtex4}
\usepackage{mathrsfs}

\usepackage{graphicx}

\begin{document}

\title{Doping Evolution of Nodal Band Renormalization in Bi$_2$Sr$_2$CuO$_{6+\delta}$
Superconductor Revealed by Laser-Based Angle-Resolved Photoemission Spectroscopy}
\author{Yingying Peng$^{1}$, Jianqiao Meng$^{1}$, Lin Zhao$^{1}$,
Yan Liu$^{1}$, Junfeng He$^{1}$, Guodong Liu$^{1}$,
Xiaoli Dong$^{1}$, Shaolong He$^{1}$, Jun Zhang$^{1}$,
Chuangtian Chen$^{2}$, Zuyan Xu$^{2}$ and X. J. Zhou$^{1,*}$}

\affiliation{
\\$^{1}$National Laboratory for Superconductivity, Beijing National Laboratory for Condensed
Matter Physics, Institute of Physics, Chinese Academy of Sciences,
Beijing 100190, China
\\$^{2}$Technical Institute of Physics and Chemistry, Chinese Academy of Sciences, Beijing 100190, China}

\date{May 5, 2013}

\begin{abstract}

{\bf
High resolution laser-based angle-resolved photoemission measurements have been carried out
on Bi$_2$Sr$_2$CuO$_{6+\delta}$ superconductor covering a wide doping range from heavily
underdoped to heavily overdoped samples. Two obvious energy scales are identified in the nodal
dispersions: one is the well-known 50-80 meV high energy kink and the other is $<$10 meV low
energy kink. The high energy kink increases monotonously in its energy scale with increasing
doping and shows weak temperature dependence, while the low energy kink exhibits a
non-monotonic doping dependence with its coupling strength enhanced sharply below $T_c$. These
systematic investigations on the doping and temperature dependence of these two energy scales
favor electron-phonon interactions as their origin. They point to the importance in involving
the electron-phonon coupling in understanding the physical properties and the superconductivity
mechanism of high temperature cuprate superconductors.}

\end{abstract}


\maketitle

In conventional superconductors, the electron-phonon interaction is crucial in understanding
the formation of the Cooper pairs. In high temperature cuprate superconductors, in order to
unravel the electron pairing mechanism, it is important to investigate the interactions of
electrons with other excitations, i.e., many body effects. Electron coupling with bosonic
mode leads to band renormalization that can be probed directly by angle-resolved photoemission
spectroscopy(ARPES) \cite{reviews}.  The 50$\sim$80 meV dispersion kink near the nodal
region\cite{Bogdanov,Johnson,Kaminski,Lanzara,Zhounature,ZhouPRL,Kordyuk,Meevasana,Zhang,He} and the $\sim$40 meV kink near the antinodal
region \cite{He,Gromko,Kim,Cuk} have been reported for cuprates although the nature of the
boson mode(s) involved remains under debate. Recently, a low energy kink
below 10 meV has also been reported near the nodal region
in Bi$_2$Sr$_2$CaCuO$_{8+\delta}$ (Bi2212)\cite{Zhang,Rameau,Vishik,Plumb,Anzai}
and Bi$_2$Sr$_2$CuO$_{6+\delta}$\cite{Kondo} with electron coupling with an acoustic
phonon as a possible origin\cite{Johnston}.  Systematic investigations on how the band renormalizations
evolve with doping and temperature will be helpful to understand the nature of
the bosonic mode(s) involved and the role of these many-body effects in dictating
the physical properties and superconductivity mechanism of the high temperature cuprate superconductors.

In this paper we report detailed doping and temperature dependence of the nodal
band renormalization in (Bi,Pb)$_2$(Sr,La)$_2$CuO$_{6+\delta}$ (Bi2201) by using
high resolution laser-based ARPES and covering a wide doping range from heavily
underdoped to heavily overdoped samples. We reveal two energy scales in the nodal
band dispersion, one is the well-known 50-80 meV high energy kink and the other
is $<$10 meV low energy kink. The high energy kink increases in its energy continuously
with increasing doping and shows weak temperature dependence, while the low energy kink
shows a non-monotonic doping dependence and its coupling strength increases sharply
below $T_c$.  These characteristics can be attributed to electron coupling with phonon
modes. They point to the importance in including the electron-phonon interactions in
understanding the physical properties of high temperature superconductors.

The ARPES measurements were performed on our vacuum ultraviolet(VUV) laser-based
angle-resolved photoemission system with advantages of high photon flux, enhanced bulk
sensitivity and super-high energy and momentum resolution\cite{Liu}. The photon energy
is 6.994 eV with a bandwidth of 0.26 meV and we set the energy resolution of the electron
energy analyzer (Scienta R4000) at 1 meV, giving rise to an overall energy resolution
of 1.03 meV. The angular resolution is $\sim$0.3$^{\circ}$, corresponding to a momentum
resolution $\sim$0.004${\AA}$$^{-1}$ for the 6.994 eV photon energy. The
Bi$_2$Sr$_{2-x}$La$_{x}$CuO$_{6+\delta}$ (La-Bi2201)\cite{Meng,Peng} and
the Bi$_{2-y}$Pb$_y$Sr$_2$CuO$_{6+\delta}$ (Pb-Bi2201)\cite{Zhao} single crystals were
grown by the traveling solvent floating zone method. Various doping levels are obtained
by La-doping and$\slash$or Pb-doping plus post-annealing under different conditions. For
convenience, we define hereafter the name of the samples according to their
doping (underdoped (UD), optimally-doped (OP) and overdoped (OD)) and superconducting
transition temperature $T_c$. We obtained UD3K, UD18K, UD28K and OP32K from La-doped Bi2201,
plus  OD20K and OD5K from Pb-doped Bi2201. The hole doping levels are estimated to
be 0.10 (UD3K), 0.11 (UD18K), 0.13 (UD28K), 0.16 (OP32K), 0.23 (OD20K) and 0.26 (OD5K) from an
empirical relation\cite{Presland}. The samples were cleaved \emph{in situ} in vacuum and measured
with a base pressure better than 5$\times$10$^{-11}$ Torr.

\begin{figure}[htb]
\begin{center}
\includegraphics[width=0.90\textwidth]{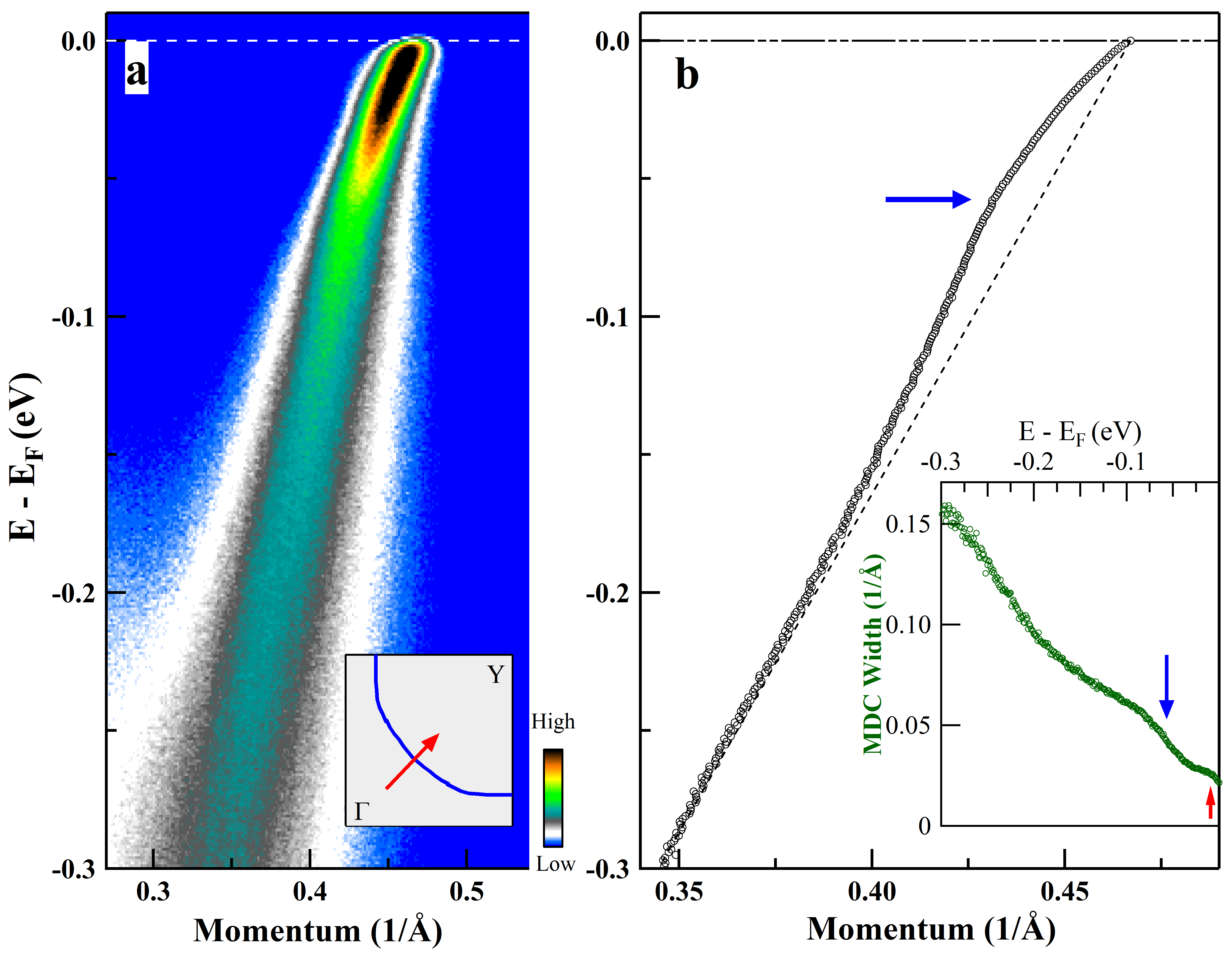}
\end{center}
\caption{Energy dispersion of underdoped
Bi2201 ($T_c$=18 K) measured along the $\Gamma(0,0)-Y(\pi,\pi)$ nodal direction at 14 K.
(a) Raw image showing photoelectron intensity as a function of energy and momentum. The
inset shows the location of the momentum cut in the Brillouin zone.  (b) Nodal dispersion
extracted from (a) by MDC fitting.  The inset in (c) shows the corresponding MDC width (FWHM).}
\end{figure}

Fig. 1a shows a typical photoemission image for an underdoped Bi2201
superconductor ($T_c$=18 K) measured along the $\Gamma(0,0)-Y(\pi,\pi)$ nodal direction at
 a temperature of 14 K. By fitting the momentum distribution curves (MDCs) at different
 binding energies, we can quantitatively obtain the dispersion (Fig. 1b) and the MDC
 width(full-width-at-half-maximum, FWHM, inset of Fig. 1b) from Fig. 1a. A prominent
 dispersion kink near $\sim$60 meV is clear from Fig. 1a and Fig. 1b, accompanied by
 the corresponding drop in the MDC width (inset of Fig. 1b). In addition, one can
 identify a low energy drop near 7 meV in the MDC width (red arrow in the inset
 of Fig. 1b). In the following, we will focus on these two energy
 scales: the 50$\sim$80 meV high energy kink and $<$10 meV low energy kink.

\begin{figure}[htb]
\begin{center}
\includegraphics[width=0.95\textwidth]{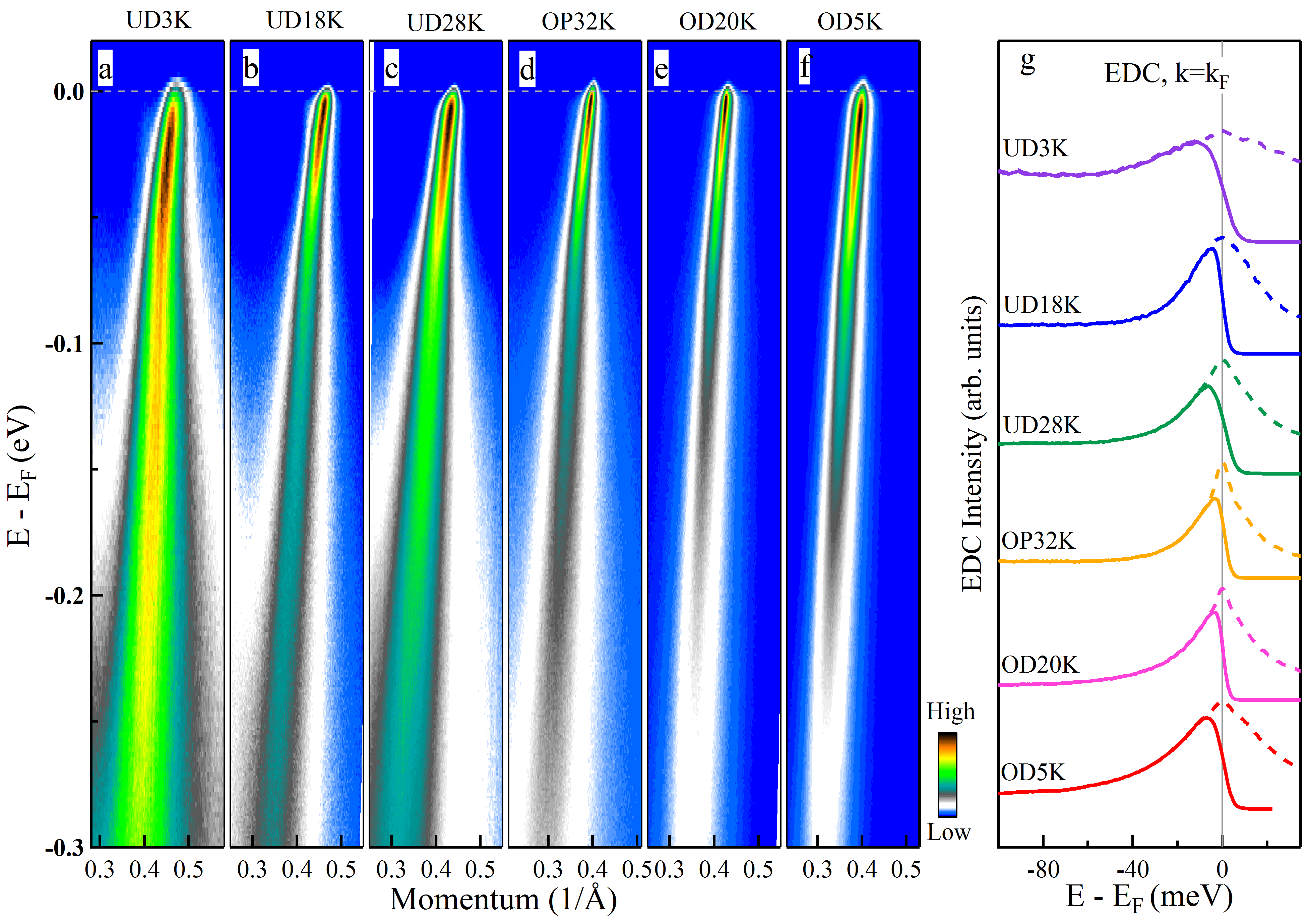}
\end{center}
\caption{Doping evolution of the nodal dispersions of Bi2201 samples measured
at 15 K. (a)-(f) Raw images of the nodal dispersions of UD3K, UD18K, UD28K,
OP32K, OD20K and OD5K, measured at 15 K, respectively.  (g) EDCs (solid lines)
and symmetrized EDCs (dashed lines) at $k_F$. For clarity, the EDCs are offset along the vertical axis.}
\end{figure}

Fig. 2 shows the doping dependence of the nodal dispersion for the Bi2201 samples.
The energy distribution curves (EDCs) at the Fermi momentum $k_F$ are shown in
Fig. 2g and these spectra indicate no gap along the nodal direction because the
symmetrized EDCs form a peak at the Fermi level\cite{Peng,Norman}. Quantitative analysis
of the dispersion and the MDC width are shown in Fig. 3.  The effective real part
of the electron self-energy, which represents the energy difference between the
measured dispersion and the empirical bare band, is shown in Fig. 3c.  The
high-energy 50$\sim$80 meV energy scale is clear in the nodal dispersion (Fig. 3a),
and manifest itself as a prominent peak in the effective real part of the electron
self-energy (Fig. 3c). The peak is broad for the underdoped samples and becomes
sharper for the overdoped samples. With increasing doping, there is a significant
increase of the peak position from $\sim$53 meV for the UD3K sample to $\sim$75 meV
for the OD5K sample, as seen from Fig. 3c and plotted in Fig. 3e (red solid circles).
Indication of the low energy kink can be observed in the expanded low energy dispersion
near the Fermi level ($E_F$)(Fig. 3b), where the dispersion within [-10 meV, $E_F$]
deviates from a linear line at higher binding energy. The low energy kink undergoes
a non-monotonic evolution with doping. For heavily underdoped sample like UD3K, the
low energy kink is weak. It gets pronounced with increasing doping, such as the
UD18K and particularly UD28K samples. The low energy kink gets weaker again with
further increase of doping, and eventually becomes nearly invisible in the heavily
overdoped samples. The energy scale of the low energy kink goes up first with doping,
reaches a maximum for the UD28K sample, and goes down with further increase of doping,
as plotted in Fig. 3e.  The doping evolution of the two energy scales also has good
correspondence in the MDC width (Fig. 3d) where one drop at high energy kink position and
the other downturn near $E_F$ can be observed. Our coverage of a wide doping range,
particularly the heavily underdoped region, made us possible to reveal a non-monotonic
doping dependence of the low energy kink; previous measurement for higher doping samples
gave only a monotonic behavior\cite{Kondo}.  Fig. 3e also plots the empirical coupling
strength ($\lambda$, black solid triangle) estimated via $\lambda=-(\partial Re\Sigma/\partial\omega)_{E_F}$ by
fitting {\it Re}$\Sigma$ in the low energy [-3meV,2meV] window with respect to $E_F$ to
a straight line.  The coupling strength decreases monotonically with increasing doping level.

\begin{figure}[htb]
\begin{center}
\includegraphics[width=0.95\textwidth]{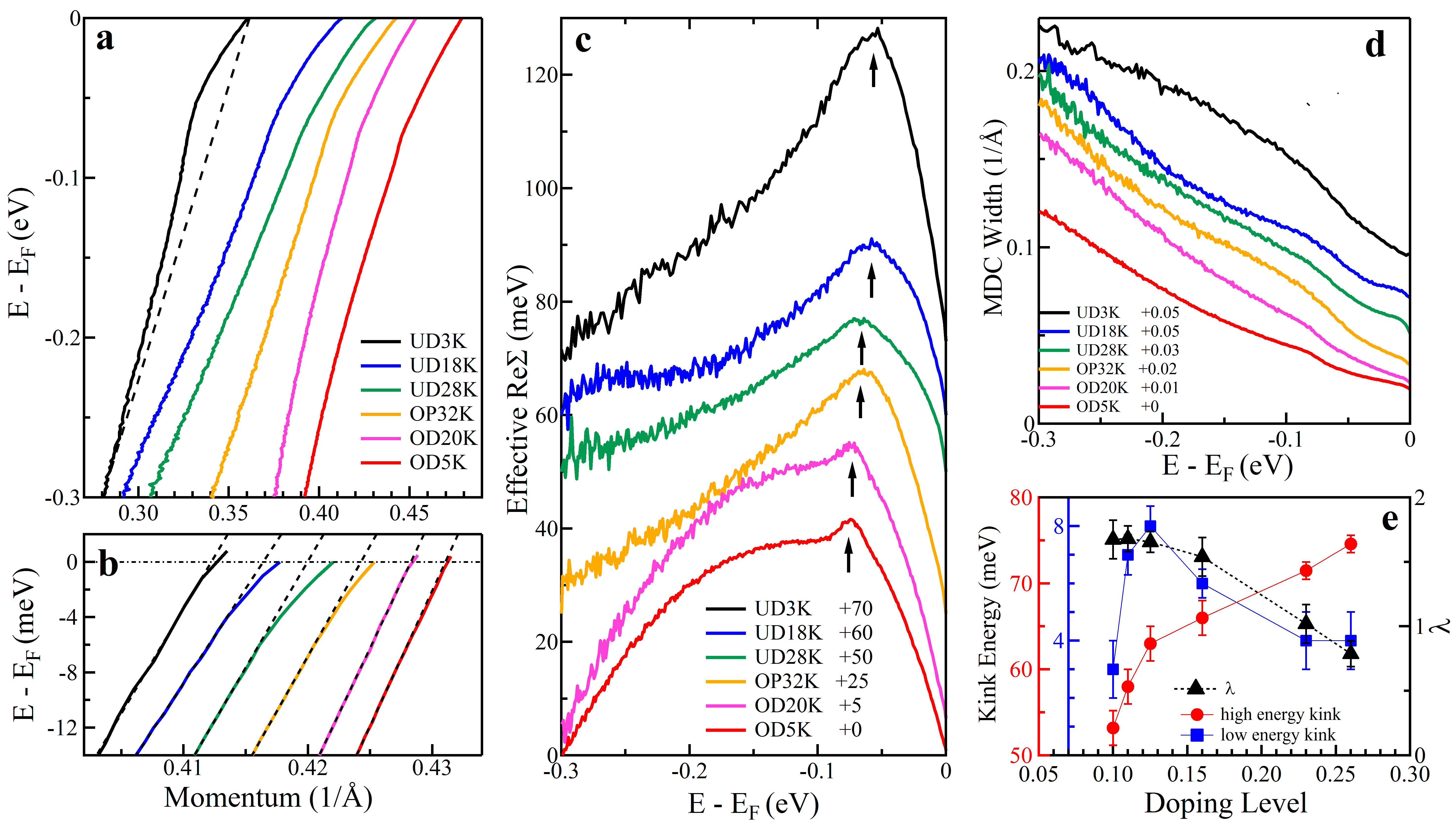}
\end{center}
\caption{Doping dependence of the nodal dispersion, self-energy and MDC width
for Bi2201 measured at 15 K. (a) MDC-derived band dispersions, offset along the
momentum axis for clarity. In order to extract the effective real part of electron
self-energy ({\it Re}$\Sigma$), an empirical bare band is assumed by a straight line
connecting the two energy positions in the dispersion at $E_F$ and -0.3 eV, as
exemplified by the dashed line accompanying the UD3K data.  (b) Same data as in
(a) expanded near the Fermi level region. Black dashed lines are linear fits
between 10-14 meV and extrapolated to $E_F$. (c) Effective {\it Re}$\Sigma$,
offset along the vertical axis for clarity with the offset number marked in the
legend. (d) Extracted MDC width (FWHM) plotted with an offset marked in legend.
(e) Doping dependence of the estimated high and low energy kink positions and the coupling constant $\lambda$. }
\end{figure}

\begin{figure}[htb]
\begin{center}
\includegraphics[width=0.95\textwidth]{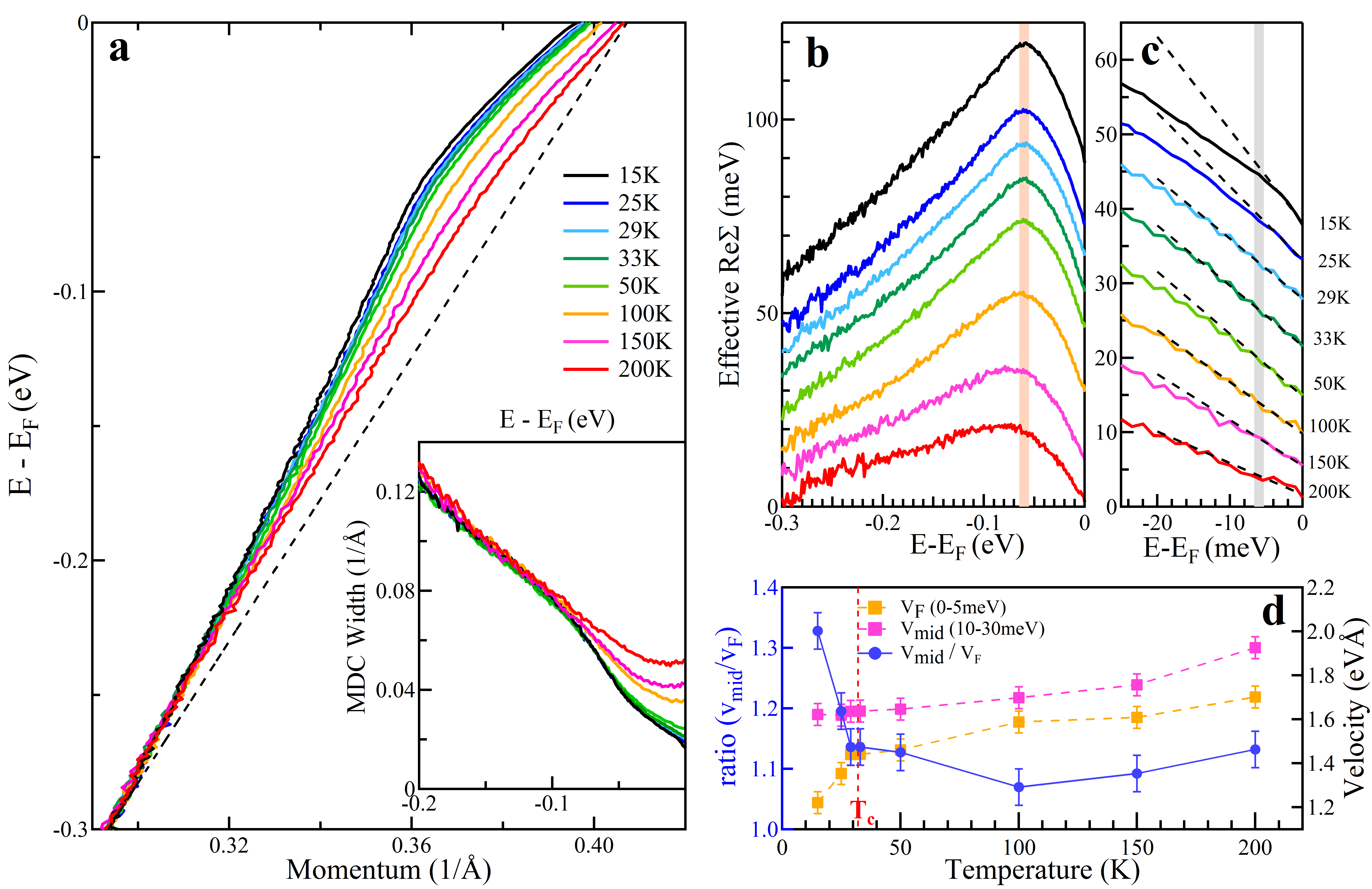}
\end{center}
\caption{Temperature dependence of the nodal electron dynamics for the OP32K
Bi2201 sample. (a) MDC-derived nodal band dispersions at different temperatures.
The corresponding MDC width is plotted in the inset.  (b) Effective {\it Re}$\Sigma$,
obtained by subtracting an empirical bare band as shown by the dashed line in (a).
The shaded bar indicates the $\sim$65 meV energy position. (c) Expanded low binding
energy region of {\it Re}$\Sigma$. Black dashed lines are fits near $E_F$. The shaded
bar indicates the location where the dispersions deviate from the linear lines ($\sim$6 meV).
(d) Temperature dependence of $v_{F}$, $v_{mid}$ and $v_{mid}/v_{F}$. The dashed red line denotes the $T_c$=32 K of the sample. }
\end{figure}

To examine the evolution of these two dispersion kinks with temperature, Fig. 4a
shows the nodal dispersions of the OP32K Bi2201 sample measured from 15 K to 200 K.
The corresponding effective real part of the electron self-energy is shown in Fig. 4b and Fig. 4c for
the expanded region near $E_F$.  From Fig. 4a, it is clear that the dispersion above 250 meV binding
energy shows little change with temperature, but the Fermi momentum at $E_F$ shifts monotonously
with temperature. Similar behavior was also observed in Bi2212\cite{Zhang}.  The obvious feature
near 65 meV gets weaker with increasing temperature, as seen from the dispersion (Fig. 4a),
the real part of the electron self-energy (Fig. 4b), as well as  the corresponding drop in
the MDC width (inset of Fig. 4a). The energy position of the $\sim$65 meV peak in the
effective {\it Re}$\Sigma$ (Fig. 4b) shows weak temperature dependence from 15 K to 200 K,
especially when crossing $T_c$.   From the expanded low binding energy portion of {\it Re}$\Sigma$
in Fig. 4c, it is clear that the {\it Re}$\Sigma$ deviates obviously from the linear fits
near $E_F$ at $\sim$6 meV at 15 K;  this deviation gets weaker with increasing temperature,
consistent with the MDC width data (inset of Fig. 4a). To illustrate the temperature evolution
of the low energy kink, we defined two velocities, $v_{mid}$ and $v_{F}$, from the linear
fit to the dispersion between 10-30 meV and 0-5 meV binding energy windows, respectively
(Fig. 4d). We can see that $v_{mid}$ shows a smooth variation across $T_c$ while $v_{F}$
shows a sudden drop below $T_c$. The low energy kink strength, which is represented by
the ratio $v_{mid}/v_{F}$, increases sharply below $T_c$. Similar results were also reported in  Bi2212\cite{Plumb}.

It is generally agreed that the 50-80 meV high energy nodal dispersion kink originates
from a coupling of electrons with a collective boson mode. However, it remains under
debate whether the boson mode is phonon\cite{Lanzara,Zhounature,Meevasana} or magnetic resonance mode\cite{Johnson,Kaminski,Kordyuk}.
The observation of such a pronounced feature in Bi2201 makes it unlikely to attribute the
bosonic mode to the magnetic resonance mode. First, considering that the magnetic resonance
mode energy is proportional to $T_c$\cite{HeH}, for the Bi2201 system with much lower $T_c$,
one would expect a much lower mode energy. For the optimally-doped Bi2201 with a $T_c$=32 K,
it is estimated to have a magnetic resonance mode with an energy of $\sim$13 meV (for optimally-doped
Bi2212 with a $T_c$$\sim$90 K, it has a resonance mode energy of $\sim$40 meV). Such a small
energy scale is not compatible with the 50-80 meV nodal kink. Second, the resonance mode is
observed only below $T_c$ for the optimally-doped and over-doped samples, but the 50-80 meV
feature is present in the normal state for the optimally-doped and overdoped Bi2201 samples
(Figs. 3c and 4b). Third, for the heavily overdoped Bi2201, the $T_c$ is rather low and the
resonance mode, if exists, must be very weak. But the 50-80 meV nodal dispersion kink is
still pronounced.  On the other hand, the present observations on the doping and temperature
dependence of the 50$\sim$80 meV energy scale are consistent with the electron coupling with
phonon mode(s) because phonons are present in materials at different dopings and at different
temperatures. In particular, the energy increase of the high energy kink with increasing doping
is also consistent with the Raman scattering result on Bi2201 where one phonon mode near $\sim$70 meV
increases in its energy with increasing doping\cite{Sugai}. In this sense, Bi2201 provides an ideal
system to unravel the origin of the 50-80 meV feature because the energy scale of the related
phonon mode and the magnetic mode are well separated in Bi2201. However, we note that more work
is needed to understand the peculiar temperature dependence of the 50$\sim$80 meV high energy
kink in terms of the electron-phonon coupling. There is no obvious energy shift of the peak in
the real part of the self-energy observed when the temperature crosses $T_c$=32 K (Fig. 4b).
Similar behavior was also observed in Bi2212\cite{Zhang,He}. In a conventional picture of electron-boson
coupling, when there is a mode with an energy $\Omega$ in the normal state, one would expect to
see a mode energy shift to $\Omega$+$\Delta$ in the superconducting state with $\Delta$ being
the superconducting gap\cite{Sandvik}. As the maximum gap of the optimally-doped Bi2201 ($T_c$=32 K)
is $\sim$15 meV\cite{Menggap},  one would expect a peak shift to higher binding energy by 15 meV
below $T_c$ which is not observed in the present measurement.

The $<$10 meV low energy kink behaves differently from the 50$\sim$80 meV high energy kink
in a couple of aspects. The first is its non-monotonic doping dependence (Fig. 3e) and the
second is its temperature sensitivity across {\it $T_c$} (Fig. 4d). Obviously the conventional
electron-boson coupling picture cannot explain the behavior of the low energy kink\cite{Sandvik}.
Since in the superconducting state, the initial mode $\Omega$ in the normal state will be
shifted to a new position  $\Omega$+$\Delta$, this means no signature of mode should exist
between $\Delta$ and $E_F$.   A recent theoretical work interpreted the low energy kink
as due to electron coupling with the in-plane polarized acoustic phonon branch:  the direct
interplay between the electron-phonon and extended Coulomb interactions would produce
forward scattering from {\it k} to {\it k+q} with small {\it q} when metallic screening
broke down in the underdoping regime\cite{Johnston}. This would result in an energy position
in the superconducting state that is determined by the initial mode energy plus a local
superconducting gap that is zero along the nodal direction.
Our observations of significant doping dependence of the low energy kink and the increase
of the coupling strength with reduced doping are in agreement with this picture. But the
weakening of the low energy kink in the heavily underdoped samples seems to be not consistent
with this picture.  It was pointed out that this low energy coupling would enhance
superconductivity as being beneficial to the d-wave pairing\cite{Johnston}. Our observation
that the low energy kink changes its coupling strength dramatically across $T_c$ also
indicates that it is closely related to superconductivity. The weak coupling strength
in the heavily underdoped samples (UD18K and UD3K) could be partly due to their relatively
low $T_c$ when compared with the measurement temperature (15 K).  According to this
theoretical scenario\cite{Johnston}, with the local gap near the nodal point is zero, the
measured low energy kink position then represents the energy of the acoustic phonon mode
involved.  Whether the acoustic phonon mode undergoes such a non-monotonic doping
variation can be examined by further experiments to directly measure the acoustic phonons.

In summary, from our high-precision ARPES measurements on Bi2201 with a wide doing range,
we revealed two energy scales in the nodal dispersion that behave differently in their
doping and temperature dependence. Our observations support that the high energy kink is
related to electron coupling with the optical phonon while the low energy kink is likely
caused by coupling to an acoustic phonon. These results reaffirm the importance of
including electron-phonon interactions  in understanding the physical properties and
superconductivity mechanism of high temperature cuprate superconductors.

This work is supported by the NSFC (Grant No. 11190022) and the MOST of China (Program No: 2011CB921703 and 2011CB605903).

$^{*}$Corresponding author: XJZhou@aphy.iphy.ac.cn
%sunnypyy@gmail.com

\end{document}